\documentclass[%
 reprint,
 amsmath,amssymb,
 aps,
prb,
]{revtex4-2}
\usepackage{filecontents}
\usepackage{graphicx}
\usepackage{dcolumn}
\usepackage{bm}
\usepackage{physics}
\usepackage{xspace}
\usepackage{times}
\usepackage{amsmath}
\usepackage{fouriernc}
\usepackage{latexsym}
\usepackage{physics}
\usepackage{xcolor}
\usepackage{mathrsfs}
\usepackage{gensymb}
\newcommand{\tn}{$T_\mathrm{N}$\xspace}
\newcommand{\etal}{\emph{et al.}}
\newcommand{\ra}{RA-SHG\xspace}

\begin{document}

\preprint{APS/123-QED}

\title{Study of the symmetry of electronic system of the Weyl semimetal GdBiPt using rotational anisotropy of second harmonic generation}
\author{S. Daneau}
\affiliation{Institut Courtois, Université de Montréal, Montréal (Québec), H2V 0B3, Canada}
\affiliation{D\'epartement de Physique, Universit\'e de Montr\'eal, Montr\'eal (Québec), H2V 0B3, Canada}
\altaffiliation{Regroupement Qu\'eb\'ecois sur les Mat\'eriaux de Pointe (RQMP), Université de Montréal, Montréal (Québec), H2V 0B3, Canada}
\author{R. Leonelli}
\affiliation{Institut Courtois, Université de Montréal, Montréal (Québec), H2V 0B3, Canada}
\affiliation{D\'epartement de Physique, Universit\'e de Montr\'eal, Montr\'eal (Québec), H2V 0B3, Canada}
\altaffiliation{Regroupement Qu\'eb\'ecois sur les Mat\'eriaux de Pointe (RQMP), Université de Montréal, Montréal (Québec), H2V 0B3, Canada}
\author{A. D. Bianchi}
\affiliation{Institut Courtois, Université de Montréal, Montréal (Québec), H2V 0B3, Canada}
\affiliation{D\'epartement de Physique, Universit\'e de Montr\'eal, Montr\'eal (Québec), H2V 0B3, Canada}
\altaffiliation{Regroupement Qu\'eb\'ecois sur les Mat\'eriaux de Pointe (RQMP), Université de Montréal, Montréal (Québec), H2V 0B3, Canada}

\date{\today}

\begin{abstract}

The half-Heusler compound GdBiPt orders antiferromagnetically around $T_\mathrm{N}=8.5$ K which implies  breaking  the time reversal symmetry as well as the translational symmetry. This combination  preserves the global symmetry. Here, we used rotational anisotropy of second harmonic generation (SHG) to study the symmetry changes associated with this phase transition. GdBiPt crystallizes in the space group $F\overline{4}3m$, which does not change through the phase transition. From powder neutron diffraction, the proposed magnetic point group for the magnetic unit cell is $3m$. We carried out a symmetry analysis of the SHG patterns. Above \tn, the SHG data shows a $C_2$ symmetry which excludes the point group  $\overline{4}3m$, as well as the $3m$ point group which represents a $[1 1 1]$ facet of the sample. Thus, we considered the point groups $2$ and $m$, but we have to reject the first one as the associated tensor elements poorly represent the angular dependence SHG. Finally, a  superposition consisting of the point groups $3m$ and $m$ fit the data correctly. Below \tn, we had to add a third contribution associated with  magnetic points group $m'$ in order to properly describe the SHG signal. The SHG intensity is linearly proportional to the antiferromagnetic order parameter. This allows us to determine $T_\mathrm{N}=9.61\pm0.48$ K and a critical exponent $\beta=0.346\pm0.017$, which are in agreement with the values found in the literature.

\end{abstract}




\maketitle


Weyl semimetals represent a fascinating frontier in quantum materials research, where the interplay of strong electronic correlations, spin-orbit coupling, and space-group symmetry gives rise to novel topological states of matter. These materials offer a unique platform for exploring fundamental physics and hold promise for groundbreaking applications in quantum-driven spintronics and quantum computing \cite{nakajima_topological_2015-1}. Here we study GdBiPt which is a half Heusler compound proposed to be a Weyl semi-metal.

Heusler compounds are magnetic intermetallics with a face-centered cubic structure, with the XYZ (half-Heuslers) or X$_2$YZ (full Heuslers) composition. In these compounds, X and Y are transition metals, while Z is a $p$-block element. Many Heusler compounds display properties critical to spintronics, including magnetoresistance, variations in the Hall effect, as well as ferro-, and antiferromagnetism \cite{wollmann_heusler_2017}. Some also exhibit superconductivity and topologically non-trivial band structures \cite{nakajima_topological_2015-1,yan_topological_2017}. Their magnetism is driven by a double exchange mechanism between neighboring magnetic ions, often involving ions in different oxidation states. The discovery of three-dimensional topological states in compounds like Bi$_2$Se$_3$ which presents an insulating gap in the crystal volume but topologically protected conductive states on the surfaces or edges have opened the way for new research in the physics of condensed matter. However, Heusler and half-Heusler compounds have the potential to be topological insulators with tunable electronic properties.

The GdBiPt crystal structure has the space group $F\overline{4}3m$ (\#216). This structure consists of four interpenetrating $fcc$ lattices shifted by $[\frac{1}{4},\frac{1}{4},\frac{1}{4}]$, three of them occupied by a different element while the fourth forms an ordered vacancy, and, more importantly, leads  to the loss of the inversion symmetry. This gives rise to strong-spin orbit coupling,  so that  the spin-degeneracy of the band structure is lifted. GdBiPt first raise to prominence was when Mong \etal,~\cite{PhysRevB.81.245209} proposed that GdBiPt may be the first antiferromagnetic (AFM) topological insulator, where both time reversal symmetry $\Theta$ and lattice translational symmetry $T_{1/2}$ are broken. However, the product $S=\Theta T_{1/2}$ is preserved. This distinguishes Heusler compounds from ordinary topological insulators.

The massless chiral (Weyl) fermions in a Weyl semimetals are expected to have a non-trivial energy dispersion. As a consequence, Weyl fermions have been predicted to present  the chiral or Adler–Bell–Jackiw anomaly, a chirality imbalance in the presence of parallel magnetic, and electric fields, leading to a large  negative longitudinal magnetoresistance (LMR)~\cite{reis_search_2016}. Such a LMR was found in GdBiPt together with the field-steering properties specific to the chiral anomaly~\cite{hirschberger_chiral_2016}. Furthermore, the chiral anomaly also induced strong suppression of the thermopower~\cite{hirschberger_chiral_2016}. GdBiPt shows an anomalous Hall angle, which is significantly larger than the ones normally observed in antiferromagnets. It is comparable in size to the largest observed in bulk ferromagnets. Neutron diffraction experiments~\cite{PhysRevB.90.041109}, and electronic structure calculations lead to the suggestion that this effect originates from Weyl points in the bandstructure~\cite{suzuki_large_2016}. A follow-up study reported  that the electrical and thermal magnetotransport in GdPtBi cannot be solely explained by Weyl physics~\cite{schindler_anisotropic_2020}. They reported that it they are strongly influenced by the interaction of the itinerant charge carriers and phonons with localized magnetic Gd ions, as well as paramagnetic impurities~\cite{schindler_anisotropic_2020}.

The development of Weyl points in the electronic structure of a material depends on the symmetry of the underlying crystal structure. The rotational anisotropy of the second harmonic generation (RA-SHG) is a nonlinear optical technique that is exquisitely sensitive to symmetry changes in the electronic system. It has been used to detect subtle distortion in the crystal structure of Sr$_2$IrO$_4$, which arises from a staggered tetragonal distortion of the oxygen octahedra~\cite{PhysRevLett.114.096404}. In the high temperature superconductor YBa$_2$Cu$_3$O$_y$ RA-SHG showed that below the pseudo-gap temperature $T^\star$ the spatial inversion and two-fold rotational symmetries are broken, while mirror symmetries perpendicular to the Cu-O plane are absent at all temperatures~\cite{zhao_global_2016}. RA-SHG is also able to detect the presence of an electronic nematic, such as observed in the spin-orbit coupled  superconductor Cd$_2$Re$_2$O$_7$~\cite{Harter295}. In topological insulators, RA-SHG was used to study the tunable surface electrons~\cite{PhysRevLett.106.057401, PhysRevB.86.035327}.

In antiferromagnets, \ra was used to study the phase transition due to its sensitivity to symmetry changes, particularly the time-reversal symmetry \cite{10.5555/1817101, butcher_cotter_1990, Jonsson25333, ShenBook, doi:10.1142/3046}. In centrosymmetric materials, SHG is forbidden in the electric dipole approximation unless the symmetry is broken. Below the Néel temperature \tn, the AFM order breaks the time-reversal symmetry, enabling SHG to detect a new response associated with the magnetic order, which is absent in the paramagnetic phase. The SHG intensity is often proportional to the odd powers of the AFM order parameter, making it a reliable method for tracking the evolution of magnetic ordering as the system transitions through \tn. In this investigation, we explore the rotational anisotropy of SHG (RA-SHG) in GdBiPt to characterize the AFM phase transition and to precisely define the associated order parameter.

\begin{figure}
\includegraphics[width=1\linewidth]{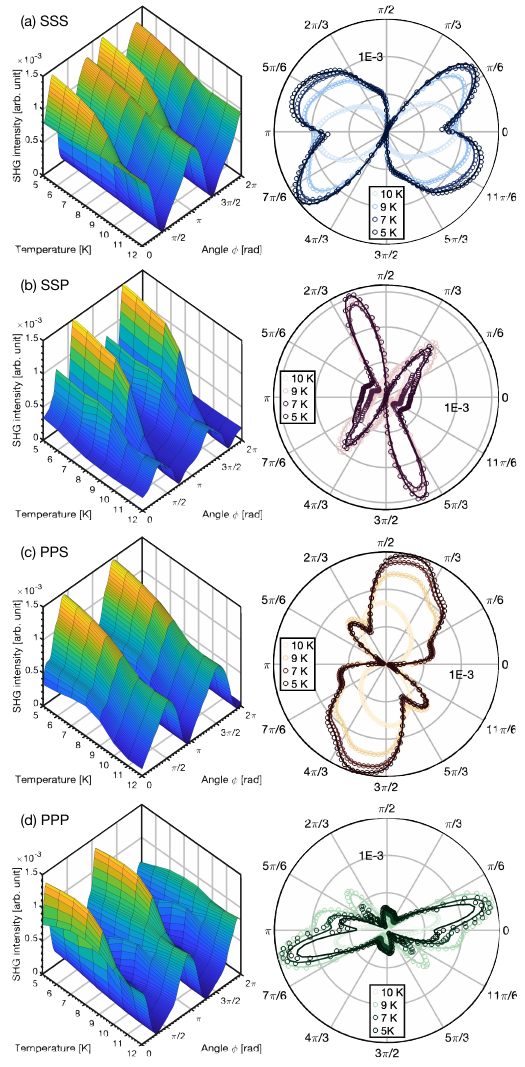}
\caption{$\vert$ \textbf{Anisotropy patterns of the SHG of GdBiPt between $12K$ and $5K$.}
Evolution of the anisotropy patterns of the different polarization contributions between $12K$ and $5K$ of GdBiPt. Fitting curves representing the coherent superposition of electric dipole emissions, $3m(i)$, $m(i)$ and $m'(c)$ are presented on top of some of the experimental data.}
\label{fig:AnisotropyPattern}
\end{figure}

\begin{figure*}
\includegraphics[width=1\linewidth]{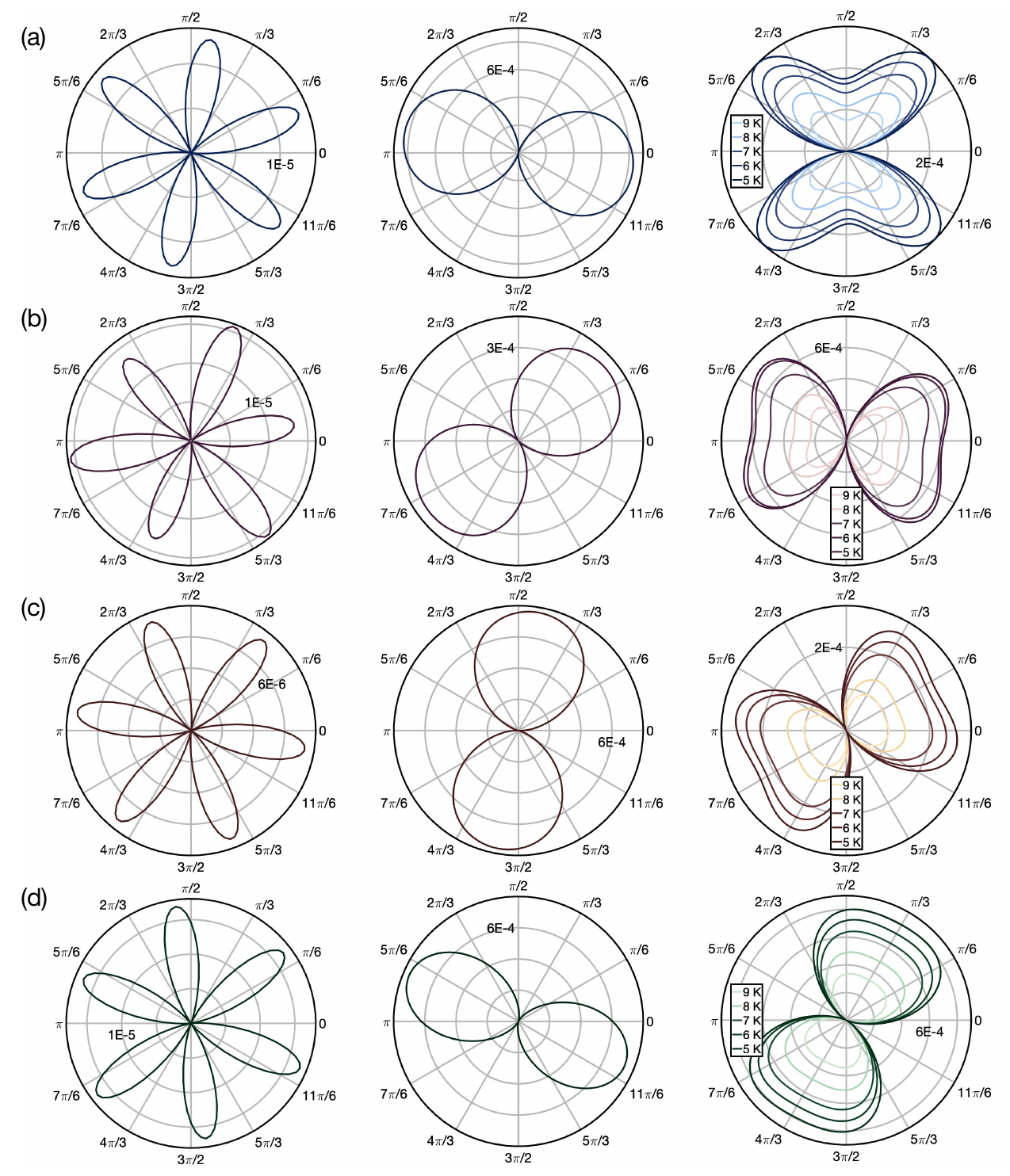}
\caption{$\vert$ \textbf{Coherent superposition
of the three electric dipolar contributions of the SHG intensity of GdBiPt.}
Each lines represent different polarization geometries: (a) SSS, (b) SSP, (c) PPS and (d) PPP. Both first columns illustrate the emission of the type-i tensors of the groups of points $3m$ and $m$, unlike the type-c tensor of the group of points $m'$ allowed below \tn in the third column. The curves respectively show $|\chi^{ED(i):\,3m}|^2$, $|\chi^{ED(i):\,m}|^2$ and $|\chi^{ED(c):\,m'}|^2$. Here, we omit the cross terms from the equation \ref{eq:GdBiPt_Pol}.}
\label{fig:CoherentSuperposition}
\end{figure*}

To analyze the \ra we need an understanding on how susceptibility tensors come about. In general, susceptibility tensors have intrinsic symmetries that together form the point group that characterize the physical properties of the medium. This postulate, known as Neumann's principle, is fundamental in the study of systems showing broken symmetries. We can express the invariance of the symmetry operations on the susceptibility tensors according to,
\begin{align}
    \chi_{\alpha\beta_1\beta_2}^{(2)} = \sum_{ab_1b_2}R_{\alpha a}R_{\beta_1b_1}R_{\beta_2b_2}\chi_{ab_1b_2}^{(2)},
\label{eq:Neumann}
\end{align}
Figure \ref{fig:AnisotropyPattern} shows the temperature-dependent evolution of RA-SHG pattern in GdBiPt for four different polarization geometries near \tn. The labelling is defined by the polarization of the two incident $\omega$ photons followed by the polarization of the measured $2\omega$ photon (SSS, SSP, PPS and PPP). Above \tn, the experimental data reveal a constant two-lobe anisotropy. This suggests that the point group $\overline{4}3m$ has too high a symmetry to account for the observed data, as it would result in six-lobe patterns. The subgroups of $\overline{4}3m$, in descending order of symmetry, include $23$, $\overline{4}2m$, $3m$, $\overline{4}$, $mm2$, $222$, $3$, $2$, $m$, and $1$. All these subgroups are non-centrosymmetric and permit SHG through electric dipole emission. Notably, the point groups $23$ and $\overline{4}2m$ belong to cubic and tetragonal crystal systems, respectively, which do not align with the atomic structure of GdBiPt along the [1 1 1] direction. However, the point group $3m$, which is trigonal, provides a more plausible explanation for the observed anisotropy in our data. 

According to \cite{0444101446,Gallego:lk5043} and with the help of the generating matrices in equation \ref{eq:Neumann}, the rank three polar tensor associated with the crystallographic electric dipole emission of the volume contains non-zero independent elements ($xxy=xyx=yxx=-yyy$, $xxz=xzx=yyz=yzy$, $zxx=zyy$, $zzz$). The transformed matrix $\chi_{\alpha\beta_1\beta_2}^{ED:\,3m}$ in the coordinate system of the experiment, used to derive the intensity of the anisotropy of SHG as a function of $\phi$, gives rise to an intensity evolution $\propto \cos{3\phi}$ for the SSS emission. Thus, we would expect a six-lobe pattern of equal amplitude. This is obviously not what is observed. Furthermore, no transformation which would include a reorientation of the illuminated face makes it possible to obtain the measured $C_2$ symmetry \cite{PhysicalReviewBVol49No20}. At the same time, the anisotropy patterns do not show perfect $C_2$ symmetry either. We see that there is always an asymmetry with respect to an axis parallel to the lobes. Nevertheless, the point groups $2$ or $m$ could be good candidates to describe the anisotropy. 

The modified tensors that account for the anisotropy of the point groups $2$ and $m$ are quite similar. To distinguish between them, a detailed analysis of the tensor elements is necessary, particularly focusing on the PPP contribution, which includes the largest number of terms. In this analysis, we observe that the PPP contribution for point group $2$ contains a term proportional to $\cos{2\phi}$, which does not seem relevant here. Specifically, the corresponding fitting coefficient becomes zero when attempting to match the data to the expected curve for point group $2$. However, this term is absent in the PPP contribution from point group $m$. Therefore, we favor point group $m$ over $2$. The non-zero independent elements for point group $m$ are ($xxx$, $xxz = xzx$, $xyy$, $xzz$, $yxy = yyx$, $yyz = yzy$, $zxx$, $zxz = zzx$, $zyy$, $zzz$). Based on these theoretical predictions, we can attempt to reconstruct the experimental data. By considering the anisotropy associated with point group $m$ alone, clear differences emerge between the experimental data and the theoretical curves. The theoretical curves exhibit mirror-plane symmetry, whereas our experimental data do not fully align with this symmetry. Although point group $m$ seems appropriate overall, the asymmetry of the lobes is not well captured by this group alone. Contributions from electric quadrupole or magnetic dipole terms are not considered, given their low intensities. Therefore, an additional component must be introduced to the electric dipole emission from the point group $m$. Through interference, this second component would account for the observed asymmetry.

It turns out that point group $3m$ provides the missing element needed to reconcile the differences. The second column of figure \ref{fig:AnisotropyPattern} illustrates the theoretical coherent superposition of electric dipole emissions from point groups $3m$ and $m$ across all polarization geometries above \tn. By comparing the amplitudes of the respective tensor elements, we find that the contribution from the point group $3m$ (on the order of $10^{-5}$) is at least an order of magnitude smaller than that from point group $m$ (on the order of $10^{-4}$). However, constructive interference amplifies its effect, significantly modifying the theoretical curves. Finally, the small contribution from point group $3m$ is consistent with our Laue backscattering pattern. The laser interaction region has a relatively low hexagonal structure factor, along with multiple reconstructions and/or crystal domains, resulting in SHG emission predominantly described by point group $m$. 

In the following, we are going to look at the \ra signal below \tn.
The literature suggests that in GdBiPt two types of AFM structures are possible below \tn \cite{PhysRevB.91.235128}. Type-A involves a spin ordering on Gd atoms, similar to the magnetic structure of CeBiPt \cite{Wosnitza_2006}, where ferromagnetic planes are aligned antiferromagnetically along the $(100)$ direction. In contrast, Type-G consists of ferromagnetic planes aligned perpendicular to the $(1 1 1)$ direction. Both structures lower the crystal symmetry due to their spin ordering. Type-A corresponds to the magnetic space group $P\overline{4}2m$, while Type-G is characterized by the magnetic space group $R3m$. As the crystallographic symmetry obtained above \tn is mainly governed by the group of points $m$, and there is no structural transition reported below \tn in the literature, we also consider a magnetic point group $m'$ to describe observable symmetry changes.  The non-zero independent elements for magnetic point group $m'$ are ($xxy = xyx$, $xyz=xzy$, $yxx$, $yxz=yzx$, $yyy$, $yzz$, $zxy=zyx$, $zyz=zzy$). Therefore, this magnetic contribution can be added to the coherent superposition above \tn and we found that it properly describe the new symmetries appearing in the AFM phase. 

To recap, the electric dipolar SHG in magnetically ordered crystals must contain a time-invariant term (i-type) which is sensitive to the spinless crystallographic lattice and a second non-time-invariant term (c-type) which is sensitive to the lattice of spin. In other words, the magnetic structure interferes with the crystallographic emission, constituting the total SHG signal \cite{Fiebig:05}. For materials with magnetic order, the relationship between the fundamental electric fields, incident on the sample, and the nonlinear polarization induced in the electric dipole approximation is given by the equation \ref{eq:GdBiPtCoherentSuperposition} which expresses the intensity of SHG. Subsequent to the results presented in the previous subsection, the expression for the induced polarization is formed from the three contributions, $\chi_{\alpha\beta_1\beta_2}^{ED(i):\,3m}$, $\chi_{\alpha\beta_1\beta_2}^{ED(i):\,m}$ and $\chi_{\alpha\beta_1\beta_2}^{ED(c):\,m'}$,
\begin{align}
\begin{split}
    \bm{P}_\alpha^{(2\omega)}(\bm{\kappa};z) & \equiv \bm{P}_\alpha^{(2\omega)}(\bm{r})e^{-2i\bm{\kappa}\cdot\bm{R}} \\
    & = \sum_{\beta_1\beta_2}\big[\chi_{\alpha\beta_1\beta_2}^{ED(i):\,3m}+\chi_{\alpha\beta_1\beta_2}^{ED(i):\,m}+\chi_{\alpha\beta_1\beta_2}^{ED(c):\,m'}\big] \\
    & \hspace{5mm} \times \bm{E}_{\beta_1}(\bm{r})\bm{E}_{\beta_2}(\bm{r})e^{-2i\bm{\kappa}\cdot\bm{R}}.
\label{eq:GdBiPt_Pol}
\end{split}
\end{align}
Thus, the polar, time-invariant tensors $\chi_{\alpha\beta_1\beta_2}^{ED(i):\,3m}$ and $\chi_{\alpha\beta_1\beta_2}^{ED(i):\,m}$ represent the temperature-independent crystallographic properties of GdBiPt, indicating that no structural transition occurs. In contrast, the time-reversal symmetry-dependent polar tensor $\chi_{\alpha\beta_1\beta_2}^{ED(c):\,m'}$ corresponds to the magnetic ordering and is associated with the magnetic space group $m'$. While the sum in equation \ref{eq:GdBiPt_Pol} cannot be decomposed to individually show the three emissions due to cross-term contributions, figure \ref{fig:CoherentSuperposition} illustrates $|\chi^{ED(i):\,3m}|^2$, $|\chi^{ED(i):\,m}|^2$, and $|\chi^{ED(c):\,m'}|^2$ for all four polarization geometries. The evolution of $|\chi^{ED(c):\,m'}|^2$ with temperature is visible, while the other two components remain unchanged above $T_N$. Another important observation is that we expect the SSS and PPS patterns to be oriented perpendicular to each other, which is approximately the case for $|\chi^{ED(i):\,3m}|^2$, $|\chi^{ED(i):\,m}|^2$, and $|\chi^{ED(c):\,m'}|^2$. The same holds for the SSP and PPP patterns. Finally, there is a notable phase shift of approximately $\pi/2$ between the intensities $|\chi^{ED(i):\,m}|^2$ and $|\chi^{ED(c):\,m'}|^2$, consistent with the phase difference between type-i and type-c tensors, as shown for multipolar contributions in the article \cite{Ahn2024}.


The AFM transition in GdBiPt is a second-order phase transition, as indicated by the divergence of the derivative of magnetic susceptibility. The associated order parameter, $\Phi$, corresponds to the sublattice magnetization $M$, with $\Phi \propto |T/T_\mathrm{N} - 1|^\beta$ for $T \leq T_\mathrm{N}$. According to renormalization group theory, the critical behavior at a second-order transition depends on the system’s dimensionality, the degrees of freedom of the order parameter $n$, and the interaction range near the phase transition. For GdBiPt, a three-dimensional AFM compound of the Heisenberg type \cite{PhysRevB.92.184432}, $n = 3$, and the calculated critical exponent is $\beta = 0.369$ \cite{Wosnitza2007}. Since the intensity of the signal is proportional to the square of the order parameter ($I^{2\omega} \propto M^2$), the scaling relation becomes $I^{2\omega} \propto |T/T_\mathrm{N} - 1|^{2\beta}$. An analysis of the temperature dependence of peak intensity of the powder neutron diffraction spectrum of GdBiPt \cite{PhysRevB.84.220408} reports $T_\mathrm{N} = 8.52 \pm 0.05$ K and $\beta = 0.33 \pm 0.02$.

The order parameter of a system can be studied by measuring the variation of its susceptibilities as a function of temperature. A notable example is the investigation of magnetic phase transitions in the Weyl semimetal Co$_3$Sn$_2$S$_2$ \cite{Ahn2024}. In magnetically ordered materials, the anisotropy of SHG is expected to change below \tn because the magnetic order breaks the time-reversal symmetry. In the electric dipole approximation, the relationship between the electric field and the induced nonlinear polarization is described by the coherent superposition of two tensors \cite{Fiebig:05}. The first, $\chi^{ED(i)}$, is type-i, which remains invariant under time-reversal and accounts for the crystallographic contribution. The second, $\chi^{ED(c)}$, is type-c, which reflects the spin-dependent contribution and only becomes active below the magnetic ordering temperature \tn. Since this material exhibits finite absorption, both tensors are complex, leading to interference,
\begin{align}
    I^{2\omega}(\phi) \propto \big|\hat{e}_\alpha^{2\omega}\big[\chi_{\alpha\beta_1\beta_2}^{ED(i)}(\phi)+\chi_{\alpha\beta_1\beta_2}^{ED(c)}(\phi)\big]\hat{e}_{\beta_1}^\omega\hat{e}_{\beta_2}^\omega\big|^2(I^\omega)^2.
\label{eq:GdBiPtCoherentSuperposition}
\end{align}
It has been observed that $\chi^{ED(c)}$ exists only in the ordered phase ($T \leq T_\mathrm{N}$), and research shows that the susceptibility is linearly proportional to the AFM order parameter. This relationship has been demonstrated, for instance, in Cr$_2$O$_3$ \cite{Sa2000,PhysRevB.54.433}. In the classical Ginzburg-Landau framework for phase transitions, the terms contributing to the free energy are determined by the symmetries of both the order parameter and the crystal lattice. By extending this standard Ginzburg-Landau approach to include magnetic properties, a more general formulation of a compound’s nonlinear magneto-optical behavior can be derived \cite{PhysRevB.57.9586}. As shown in \cite{Ahn2024}, the AFM order parameter can be studied by examining the temperature dependence of the susceptibility tensor elements $\chi^{ED(c)}$, where type-i and type-c susceptibility tensors are proportional to the even and odd powers of the order parameter, respectively. Specifically, $\chi^{ED(i)} = c_1$ and $\chi^{ED(c)} = c_2 |T/T_\mathrm{N} - 1|^\beta$, since above \tn there is no sublattice magnetization. Here, $c_1$ and $c_2$ are constants. This allows us to derive the functional form of the SHG intensity: for $T > T_\mathrm{N}$, only type-i tensors contribute, and the intensity follows $I^{2\omega}_{T > T_\mathrm{N}} \propto (c_1)^2$, while for $T \leq T_\mathrm{N}$, additional contributions from the type-c tensor emerge,
\begin{align}
\begin{split}
    I^{2\omega}_{T\leq T_\mathrm{N}} & \propto \big[c_1+c_2|T/T_\mathrm{N}-1|^\beta\big]^2 \\ & = (c_1)^2+2c_1c_2|T/T_\mathrm{N}-1|^\beta+(c_2)^2|T/T_\mathrm{N}-1|^{2\beta},
\end{split}
\label{eq:OrderParameterI}
\end{align}
where generally we neglect the last term given that $c_1\gg c_2$, therefore $(c_2)^2\ll c_1c_2$ \cite{Ahn2024}. It is true that the tensor elements may be larger for $\chi^{ED(i)}$ than for $\chi^{ED(c)}$. Alternatively, if it's not the case, we define a normalized version of the equation \ref{eq:OrderParameterI},
\begin{align}
    \frac{I^{2\omega}_{T\leq T_\mathrm{N}}}{I^{2\omega}_{T>T_\mathrm{N}}} = 1+\frac{2c_2}{c_1}\big|T/T_N-1\big|^\beta+\bigg(\frac{c_2}{c_1}\bigg)^2\big|T/T_\mathrm{N}-1\big|^{2\beta}.
\label{eq:OrderParameterI/I}
\end{align}

\begin{figure}
\includegraphics[width=0.9\linewidth]{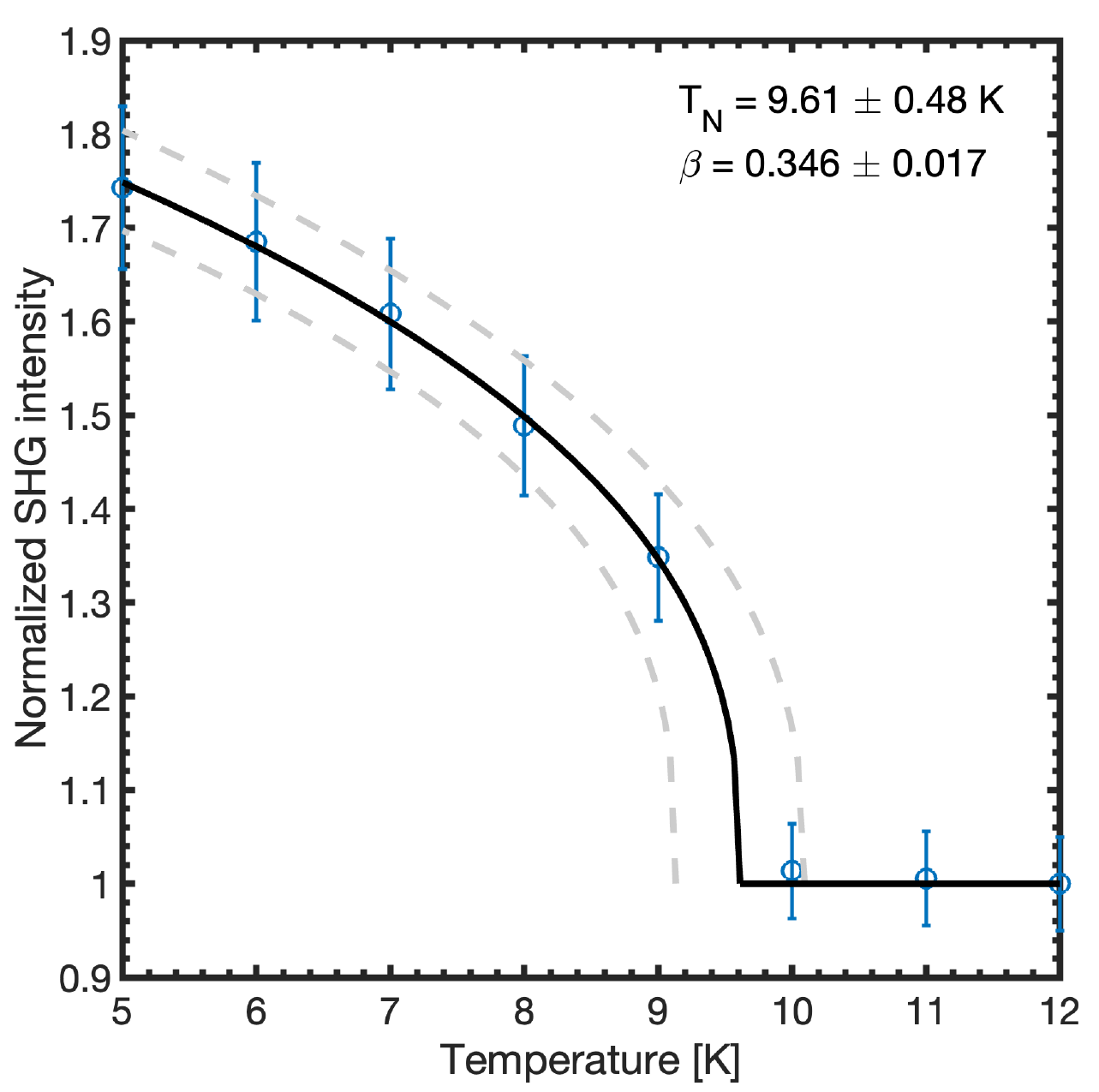}
\caption{$\vert$ \textbf{Antiferromagnetic order parameter of GdBiPt.}
In (a), SHG intensity of GdBiPt of the contribution SSS for $\phi=35\degree$ as a function of temperature. The solid line represents the equation \ref{eq:OrderParameterI/I} for which we find the following parameters: $T_\mathrm{N}=9.61\pm0.48$ K and $\beta=0.346\pm0.017$. Gray dotted lines illustrate parameters uncertainty.}
\label{fig:OrderParameter}
\end{figure}

Figure \ref{fig:OrderParameter} shows the evolution of the SHG intensity normalized by the value at $12$ K for the contribution SSS at $\phi=35 \degree$. Despite the few experimental points, the equation \ref{eq:OrderParameterI/I} seems to fit well for parameters $T_\mathrm{N}=9.61\pm0.48$ K and $\beta=0.346\pm0.017$. Thus, the Néel temperature found is not exactly equivalent to that found in the literature, but looks very good if we include the uncertainty in the actual temperature of the sample. Indeed, the thermal conductivity of GdBiPt is of the order of $\approx10$ W/Km for the temperature range of our experiment. This means that the sample quickly reaches the cold finger temperature in the cryostat and confirms an uncertainty of a few kelvins \cite{Hirschberger2016}. Moreover, this article reports \tn for GdBiPt much closer to $9$~K. As for the order parameter, the expected value for a three-dimensional AFM compound of the Heisenberg type \cite{PhysRevB.92.184432} is $\beta=0.369$. The numerical uncertainty therefore places us at the limit of this theoretical data. Nevertheless, the value of critical exponent we find is in agreement with the one found by other methods such as X-ray magnetic resonance \cite{Wosnitza2007,PhysRevB.84.220408}.

In summary, GdBiPt, a half-Heusler compound, undergoes an AFM phase transition, breaking time-reversal and translational symmetry while preserving overall symmetry. SHG anisotropy studies showed that the $C_2$ symmetry rules out the $\overline{4}3m$ point group, favoring $3m$ for the $[1 1 1]$ facet. Analysis of point groups $2$ and $m$ led to the rejection of the $2$ group due to its inadequate angular dependence. An interference combination of the $3m$ and $m$ point groups provided the best fit to the anisotropy above \tn. Below \tn, the magnetic point group $m'$ is required to describe the data, linking SHG intensity to the AFM order parameter, yielding $T_\mathrm{N} = 9.61 \pm 0.48$ K and $\beta = 0.346 \pm 0.017$, consistent with literature values of Heisenberg type AFM. The dipolar emission is described by a polar tensor $\chi^{ED(c):\,m'}$, which is compatible with magnetic order but breaks time-reversal symmetry. While more data around the AFM transition could fully confirm the Ginzburg-Landau model, dipolar emission from the $m'$ group is expected above the ordering temperature. 






\begin{acknowledgments}
The research at UdeM received support from the Natural Sciences and Engineering Research Council of Canada (Canada), the Fonds Qu\'eb\'ecois de la Recherche sur la Nature et les Technologies (Qu\'ebec), and the Canada Excellence Research Chair in Light-Matter Interaction held by Prof. Silva at UdeM.
\end{acknowledgments}


\begin{thebibliography}{33}%
\makeatletter
\providecommand \@ifxundefined [1]{%
 \@ifx{#1\undefined}
}%
\providecommand \@ifnum [1]{%
 \ifnum #1\expandafter \@firstoftwo
 \else \expandafter \@secondoftwo
 \fi
}%
\providecommand \@ifx [1]{%
 \ifx #1\expandafter \@firstoftwo
 \else \expandafter \@secondoftwo
 \fi
}%
\providecommand \natexlab [1]{#1}%
\providecommand \enquote  [1]{``#1''}%
\providecommand \bibnamefont  [1]{#1}%
\providecommand \bibfnamefont [1]{#1}%
\providecommand \citenamefont [1]{#1}%
\providecommand \href@noop [0]{\@secondoftwo}%
\providecommand \href [0]{\begingroup \@sanitize@url \@href}%
\providecommand \@href[1]{\@@startlink{#1}\@@href}%
\providecommand \@@href[1]{\endgroup#1\@@endlink}%
\providecommand \@sanitize@url [0]{\catcode `\\12\catcode `\$12\catcode `\&12\catcode `\#12\catcode `\^12\catcode `\_12\catcode `\%12\relax}%
\providecommand \@@startlink[1]{}%
\providecommand \@@endlink[0]{}%
\providecommand \url  [0]{\begingroup\@sanitize@url \@url }%
\providecommand \@url [1]{\endgroup\@href {#1}{\urlprefix }}%
\providecommand \urlprefix  [0]{URL }%
\providecommand \Eprint [0]{\href }%
\providecommand \doibase [0]{https://doi.org/}%
\providecommand \selectlanguage [0]{\@gobble}%
\providecommand \bibinfo  [0]{\@secondoftwo}%
\providecommand \bibfield  [0]{\@secondoftwo}%
\providecommand \translation [1]{[#1]}%
\providecommand \BibitemOpen [0]{}%
\providecommand \bibitemStop [0]{}%
\providecommand \bibitemNoStop [0]{.\EOS\space}%
\providecommand \EOS [0]{\spacefactor3000\relax}%
\providecommand \BibitemShut  [1]{\csname bibitem#1\endcsname}%
\let\auto@bib@innerbib\@empty
\bibitem [{\citenamefont {Nakajima}\ \emph {et~al.}(2015)\citenamefont {Nakajima}, \citenamefont {Hu}, \citenamefont {Kirshenbaum}, \citenamefont {Hughes}, \citenamefont {Syers}, \citenamefont {Wang}, \citenamefont {Wang}, \citenamefont {Wang}, \citenamefont {Saha}, \citenamefont {Pratt}, \citenamefont {Lynn},\ and\ \citenamefont {Paglione}}]{nakajima_topological_2015-1}%
  \BibitemOpen
  \bibfield  {author} {\bibinfo {author} {\bibfnamefont {Y.}~\bibnamefont {Nakajima}}, \bibinfo {author} {\bibfnamefont {R.}~\bibnamefont {Hu}}, \bibinfo {author} {\bibfnamefont {K.}~\bibnamefont {Kirshenbaum}}, \bibinfo {author} {\bibfnamefont {A.}~\bibnamefont {Hughes}}, \bibinfo {author} {\bibfnamefont {P.}~\bibnamefont {Syers}}, \bibinfo {author} {\bibfnamefont {X.}~\bibnamefont {Wang}}, \bibinfo {author} {\bibfnamefont {K.}~\bibnamefont {Wang}}, \bibinfo {author} {\bibfnamefont {R.}~\bibnamefont {Wang}}, \bibinfo {author} {\bibfnamefont {S.~R.}\ \bibnamefont {Saha}}, \bibinfo {author} {\bibfnamefont {D.}~\bibnamefont {Pratt}}, \bibinfo {author} {\bibfnamefont {J.~W.}\ \bibnamefont {Lynn}},\ and\ \bibinfo {author} {\bibfnamefont {J.}~\bibnamefont {Paglione}},\ }\href {https://doi.org/10.1126/sciadv.1500242} {\bibfield  {journal} {\bibinfo  {journal} {Science Advances}\ }\textbf {\bibinfo {volume} {1}},\ \bibinfo {pages} {e1500242} (\bibinfo {year} {2015})}\BibitemShut {NoStop}%
\bibitem [{\citenamefont {Wollmann}\ \emph {et~al.}(2017)\citenamefont {Wollmann}, \citenamefont {Nayak}, \citenamefont {Parkin},\ and\ \citenamefont {Felser}}]{wollmann_heusler_2017}%
  \BibitemOpen
  \bibfield  {author} {\bibinfo {author} {\bibfnamefont {L.}~\bibnamefont {Wollmann}}, \bibinfo {author} {\bibfnamefont {A.~K.}\ \bibnamefont {Nayak}}, \bibinfo {author} {\bibfnamefont {S.~S.}\ \bibnamefont {Parkin}},\ and\ \bibinfo {author} {\bibfnamefont {C.}~\bibnamefont {Felser}},\ }\href {https://doi.org/10.1146/annurev-matsci-070616-123928} {\bibfield  {journal} {\bibinfo  {journal} {Annual Review of Materials Research}\ }\textbf {\bibinfo {volume} {47}},\ \bibinfo {pages} {247} (\bibinfo {year} {2017})}\BibitemShut {NoStop}%
\bibitem [{\citenamefont {Yan}\ and\ \citenamefont {Felser}(2017)}]{yan_topological_2017}%
  \BibitemOpen
  \bibfield  {author} {\bibinfo {author} {\bibfnamefont {B.}~\bibnamefont {Yan}}\ and\ \bibinfo {author} {\bibfnamefont {C.}~\bibnamefont {Felser}},\ }\href {https://doi.org/10.1146/annurev-conmatphys-031016-025458} {\bibfield  {journal} {\bibinfo  {journal} {Annual Review of Condensed Matter Physics}\ }\textbf {\bibinfo {volume} {8}},\ \bibinfo {pages} {337} (\bibinfo {year} {2017})}\BibitemShut {NoStop}%
\bibitem [{\citenamefont {Mong}\ \emph {et~al.}(2010)\citenamefont {Mong}, \citenamefont {Essin},\ and\ \citenamefont {Moore}}]{PhysRevB.81.245209}%
  \BibitemOpen
  \bibfield  {author} {\bibinfo {author} {\bibfnamefont {R.~S.~K.}\ \bibnamefont {Mong}}, \bibinfo {author} {\bibfnamefont {A.~M.}\ \bibnamefont {Essin}},\ and\ \bibinfo {author} {\bibfnamefont {J.~E.}\ \bibnamefont {Moore}},\ }\href {https://doi.org/10.1103/PhysRevB.81.245209} {\bibfield  {journal} {\bibinfo  {journal} {Phys. Rev. B}\ }\textbf {\bibinfo {volume} {81}},\ \bibinfo {pages} {245209} (\bibinfo {year} {2010})}\BibitemShut {NoStop}%
\bibitem [{\citenamefont {Reis}\ \emph {et~al.}(2016)\citenamefont {Reis}, \citenamefont {Ajeesh}, \citenamefont {Kumar}, \citenamefont {Arnold}, \citenamefont {Shekhar}, \citenamefont {Naumann}, \citenamefont {Schmidt}, \citenamefont {Nicklas},\ and\ \citenamefont {Hassinger}}]{reis_search_2016}%
  \BibitemOpen
  \bibfield  {author} {\bibinfo {author} {\bibfnamefont {R.~D.~d.}\ \bibnamefont {Reis}}, \bibinfo {author} {\bibfnamefont {M.~O.}\ \bibnamefont {Ajeesh}}, \bibinfo {author} {\bibfnamefont {N.}~\bibnamefont {Kumar}}, \bibinfo {author} {\bibfnamefont {F.}~\bibnamefont {Arnold}}, \bibinfo {author} {\bibfnamefont {C.}~\bibnamefont {Shekhar}}, \bibinfo {author} {\bibfnamefont {M.}~\bibnamefont {Naumann}}, \bibinfo {author} {\bibfnamefont {M.}~\bibnamefont {Schmidt}}, \bibinfo {author} {\bibfnamefont {M.}~\bibnamefont {Nicklas}},\ and\ \bibinfo {author} {\bibfnamefont {E.}~\bibnamefont {Hassinger}},\ }\href {https://doi.org/10.1088/1367-2630/18/8/085006} {\bibfield  {journal} {\bibinfo  {journal} {New Journal of Physics}\ }\textbf {\bibinfo {volume} {18}},\ \bibinfo {pages} {085006} (\bibinfo {year} {2016})}\BibitemShut {NoStop}%
\bibitem [{\citenamefont {Hirschberger}\ \emph {et~al.}(2016{\natexlab{a}})\citenamefont {Hirschberger}, \citenamefont {Kushwaha}, \citenamefont {Wang}, \citenamefont {Gibson}, \citenamefont {Liang}, \citenamefont {Belvin}, \citenamefont {Bernevig}, \citenamefont {Cava},\ and\ \citenamefont {Ong}}]{hirschberger_chiral_2016}%
  \BibitemOpen
  \bibfield  {author} {\bibinfo {author} {\bibfnamefont {M.}~\bibnamefont {Hirschberger}}, \bibinfo {author} {\bibfnamefont {S.}~\bibnamefont {Kushwaha}}, \bibinfo {author} {\bibfnamefont {Z.}~\bibnamefont {Wang}}, \bibinfo {author} {\bibfnamefont {Q.}~\bibnamefont {Gibson}}, \bibinfo {author} {\bibfnamefont {S.}~\bibnamefont {Liang}}, \bibinfo {author} {\bibfnamefont {C.~A.}\ \bibnamefont {Belvin}}, \bibinfo {author} {\bibfnamefont {B.~A.}\ \bibnamefont {Bernevig}}, \bibinfo {author} {\bibfnamefont {R.~J.}\ \bibnamefont {Cava}},\ and\ \bibinfo {author} {\bibfnamefont {N.~P.}\ \bibnamefont {Ong}},\ }\href {https://doi.org/10.1038/nmat4684} {\bibfield  {journal} {\bibinfo  {journal} {Nature Materials}\ }\textbf {\bibinfo {volume} {15}},\ \bibinfo {pages} {1161} (\bibinfo {year} {2016}{\natexlab{a}})}\BibitemShut {NoStop}%
\bibitem [{\citenamefont {M\"uller}\ \emph {et~al.}(2014)\citenamefont {M\"uller}, \citenamefont {Lee-Hone}, \citenamefont {Lapointe}, \citenamefont {Ryan}, \citenamefont {Pereg-Barnea}, \citenamefont {Bianchi}, \citenamefont {Mozharivskyj},\ and\ \citenamefont {Flacau}}]{PhysRevB.90.041109}%
  \BibitemOpen
  \bibfield  {author} {\bibinfo {author} {\bibfnamefont {R.~A.}\ \bibnamefont {M\"uller}}, \bibinfo {author} {\bibfnamefont {N.~R.}\ \bibnamefont {Lee-Hone}}, \bibinfo {author} {\bibfnamefont {L.}~\bibnamefont {Lapointe}}, \bibinfo {author} {\bibfnamefont {D.~H.}\ \bibnamefont {Ryan}}, \bibinfo {author} {\bibfnamefont {T.}~\bibnamefont {Pereg-Barnea}}, \bibinfo {author} {\bibfnamefont {A.~D.}\ \bibnamefont {Bianchi}}, \bibinfo {author} {\bibfnamefont {Y.}~\bibnamefont {Mozharivskyj}},\ and\ \bibinfo {author} {\bibfnamefont {R.}~\bibnamefont {Flacau}},\ }\href {https://doi.org/10.1103/PhysRevB.90.041109} {\bibfield  {journal} {\bibinfo  {journal} {Phys. Rev. B}\ }\textbf {\bibinfo {volume} {90}},\ \bibinfo {pages} {041109} (\bibinfo {year} {2014})}\BibitemShut {NoStop}%
\bibitem [{\citenamefont {Suzuki}\ \emph {et~al.}(2016)\citenamefont {Suzuki}, \citenamefont {Chisnell}, \citenamefont {Devarakonda}, \citenamefont {Liu}, \citenamefont {Feng}, \citenamefont {Xiao}, \citenamefont {Lynn},\ and\ \citenamefont {Checkelsky}}]{suzuki_large_2016}%
  \BibitemOpen
  \bibfield  {author} {\bibinfo {author} {\bibfnamefont {T.}~\bibnamefont {Suzuki}}, \bibinfo {author} {\bibfnamefont {R.}~\bibnamefont {Chisnell}}, \bibinfo {author} {\bibfnamefont {A.}~\bibnamefont {Devarakonda}}, \bibinfo {author} {\bibfnamefont {Y.-T.}\ \bibnamefont {Liu}}, \bibinfo {author} {\bibfnamefont {W.}~\bibnamefont {Feng}}, \bibinfo {author} {\bibfnamefont {D.}~\bibnamefont {Xiao}}, \bibinfo {author} {\bibfnamefont {J.~W.}\ \bibnamefont {Lynn}},\ and\ \bibinfo {author} {\bibfnamefont {J.~G.}\ \bibnamefont {Checkelsky}},\ }\bibfield  {journal} {\bibinfo  {journal} {Nature Physics}\ }\href {https://doi.org/10.1038/nphys3831} {10.1038/nphys3831} (\bibinfo {year} {2016})\BibitemShut {NoStop}%
\bibitem [{\citenamefont {Schindler}\ \emph {et~al.}(2020)\citenamefont {Schindler}, \citenamefont {Galeski}, \citenamefont {Schnelle}, \citenamefont {Wawrzyńczak}, \citenamefont {Abdel-Haq}, \citenamefont {Guin}, \citenamefont {Kroder}, \citenamefont {Kumar}, \citenamefont {Fu}, \citenamefont {Borrmann}, \citenamefont {Shekhar}, \citenamefont {Felser}, \citenamefont {Meng}, \citenamefont {Grushin}, \citenamefont {Zhang}, \citenamefont {Sun},\ and\ \citenamefont {Gooth}}]{schindler_anisotropic_2020}%
  \BibitemOpen
  \bibfield  {author} {\bibinfo {author} {\bibfnamefont {C.}~\bibnamefont {Schindler}}, \bibinfo {author} {\bibfnamefont {S.}~\bibnamefont {Galeski}}, \bibinfo {author} {\bibfnamefont {W.}~\bibnamefont {Schnelle}}, \bibinfo {author} {\bibfnamefont {R.}~\bibnamefont {Wawrzyńczak}}, \bibinfo {author} {\bibfnamefont {W.}~\bibnamefont {Abdel-Haq}}, \bibinfo {author} {\bibfnamefont {S.~N.}\ \bibnamefont {Guin}}, \bibinfo {author} {\bibfnamefont {J.}~\bibnamefont {Kroder}}, \bibinfo {author} {\bibfnamefont {N.}~\bibnamefont {Kumar}}, \bibinfo {author} {\bibfnamefont {C.}~\bibnamefont {Fu}}, \bibinfo {author} {\bibfnamefont {H.}~\bibnamefont {Borrmann}}, \bibinfo {author} {\bibfnamefont {C.}~\bibnamefont {Shekhar}}, \bibinfo {author} {\bibfnamefont {C.}~\bibnamefont {Felser}}, \bibinfo {author} {\bibfnamefont {T.}~\bibnamefont {Meng}}, \bibinfo {author} {\bibfnamefont {A.~G.}\ \bibnamefont {Grushin}}, \bibinfo {author} {\bibfnamefont {Y.}~\bibnamefont {Zhang}}, \bibinfo {author} {\bibfnamefont {Y.}~\bibnamefont
  {Sun}},\ and\ \bibinfo {author} {\bibfnamefont {J.}~\bibnamefont {Gooth}},\ }\href {https://doi.org/10.1103/PhysRevB.101.125119} {\bibfield  {journal} {\bibinfo  {journal} {Physical Review B}\ }\textbf {\bibinfo {volume} {101}},\ \bibinfo {pages} {125119} (\bibinfo {year} {2020})}\BibitemShut {NoStop}%
\bibitem [{\citenamefont {Torchinsky}\ \emph {et~al.}(2015)\citenamefont {Torchinsky}, \citenamefont {Chu}, \citenamefont {Zhao}, \citenamefont {Perkins}, \citenamefont {Sizyuk}, \citenamefont {Qi}, \citenamefont {Cao},\ and\ \citenamefont {Hsieh}}]{PhysRevLett.114.096404}%
  \BibitemOpen
  \bibfield  {author} {\bibinfo {author} {\bibfnamefont {D.~H.}\ \bibnamefont {Torchinsky}}, \bibinfo {author} {\bibfnamefont {H.}~\bibnamefont {Chu}}, \bibinfo {author} {\bibfnamefont {L.}~\bibnamefont {Zhao}}, \bibinfo {author} {\bibfnamefont {N.~B.}\ \bibnamefont {Perkins}}, \bibinfo {author} {\bibfnamefont {Y.}~\bibnamefont {Sizyuk}}, \bibinfo {author} {\bibfnamefont {T.}~\bibnamefont {Qi}}, \bibinfo {author} {\bibfnamefont {G.}~\bibnamefont {Cao}},\ and\ \bibinfo {author} {\bibfnamefont {D.}~\bibnamefont {Hsieh}},\ }\href {https://doi.org/10.1103/PhysRevLett.114.096404} {\bibfield  {journal} {\bibinfo  {journal} {Phys. Rev. Lett.}\ }\textbf {\bibinfo {volume} {114}},\ \bibinfo {pages} {096404} (\bibinfo {year} {2015})}\BibitemShut {NoStop}%
\bibitem [{\citenamefont {Zhao}\ \emph {et~al.}(2016)\citenamefont {Zhao}, \citenamefont {Belvin}, \citenamefont {Liang}, \citenamefont {Bonn}, \citenamefont {Hardy}, \citenamefont {Armitage},\ and\ \citenamefont {Hsieh}}]{zhao_global_2016}%
  \BibitemOpen
  \bibfield  {author} {\bibinfo {author} {\bibfnamefont {L.}~\bibnamefont {Zhao}}, \bibinfo {author} {\bibfnamefont {C.~A.}\ \bibnamefont {Belvin}}, \bibinfo {author} {\bibfnamefont {R.}~\bibnamefont {Liang}}, \bibinfo {author} {\bibfnamefont {D.~A.}\ \bibnamefont {Bonn}}, \bibinfo {author} {\bibfnamefont {W.~N.}\ \bibnamefont {Hardy}}, \bibinfo {author} {\bibfnamefont {N.~P.}\ \bibnamefont {Armitage}},\ and\ \bibinfo {author} {\bibfnamefont {D.}~\bibnamefont {Hsieh}},\ }\bibfield  {journal} {\bibinfo  {journal} {Nature Physics}\ }\href {https://doi.org/10.1038/nphys3962} {10.1038/nphys3962} (\bibinfo {year} {2016})\BibitemShut {NoStop}%
\bibitem [{\citenamefont {Harter}\ \emph {et~al.}(2017)\citenamefont {Harter}, \citenamefont {Zhao}, \citenamefont {Yan}, \citenamefont {Mandrus},\ and\ \citenamefont {Hsieh}}]{Harter295}%
  \BibitemOpen
  \bibfield  {author} {\bibinfo {author} {\bibfnamefont {J.~W.}\ \bibnamefont {Harter}}, \bibinfo {author} {\bibfnamefont {Z.~Y.}\ \bibnamefont {Zhao}}, \bibinfo {author} {\bibfnamefont {J.-Q.}\ \bibnamefont {Yan}}, \bibinfo {author} {\bibfnamefont {D.~G.}\ \bibnamefont {Mandrus}},\ and\ \bibinfo {author} {\bibfnamefont {D.}~\bibnamefont {Hsieh}},\ }\href {https://doi.org/10.1126/science.aad1188} {\bibfield  {journal} {\bibinfo  {journal} {Science}\ }\textbf {\bibinfo {volume} {356}},\ \bibinfo {pages} {295} (\bibinfo {year} {2017})}\BibitemShut {NoStop}%
\bibitem [{\citenamefont {Hsieh}\ \emph {et~al.}(2011)\citenamefont {Hsieh}, \citenamefont {McIver}, \citenamefont {Torchinsky}, \citenamefont {Gardner}, \citenamefont {Lee},\ and\ \citenamefont {Gedik}}]{PhysRevLett.106.057401}%
  \BibitemOpen
  \bibfield  {author} {\bibinfo {author} {\bibfnamefont {D.}~\bibnamefont {Hsieh}}, \bibinfo {author} {\bibfnamefont {J.~W.}\ \bibnamefont {McIver}}, \bibinfo {author} {\bibfnamefont {D.~H.}\ \bibnamefont {Torchinsky}}, \bibinfo {author} {\bibfnamefont {D.~R.}\ \bibnamefont {Gardner}}, \bibinfo {author} {\bibfnamefont {Y.~S.}\ \bibnamefont {Lee}},\ and\ \bibinfo {author} {\bibfnamefont {N.}~\bibnamefont {Gedik}},\ }\href {https://doi.org/10.1103/PhysRevLett.106.057401} {\bibfield  {journal} {\bibinfo  {journal} {Phys. Rev. Lett.}\ }\textbf {\bibinfo {volume} {106}},\ \bibinfo {pages} {057401} (\bibinfo {year} {2011})}\BibitemShut {NoStop}%
\bibitem [{\citenamefont {McIver}\ \emph {et~al.}(2012)\citenamefont {McIver}, \citenamefont {Hsieh}, \citenamefont {Drapcho}, \citenamefont {Torchinsky}, \citenamefont {Gardner}, \citenamefont {Lee},\ and\ \citenamefont {Gedik}}]{PhysRevB.86.035327}%
  \BibitemOpen
  \bibfield  {author} {\bibinfo {author} {\bibfnamefont {J.~W.}\ \bibnamefont {McIver}}, \bibinfo {author} {\bibfnamefont {D.}~\bibnamefont {Hsieh}}, \bibinfo {author} {\bibfnamefont {S.~G.}\ \bibnamefont {Drapcho}}, \bibinfo {author} {\bibfnamefont {D.~H.}\ \bibnamefont {Torchinsky}}, \bibinfo {author} {\bibfnamefont {D.~R.}\ \bibnamefont {Gardner}}, \bibinfo {author} {\bibfnamefont {Y.~S.}\ \bibnamefont {Lee}},\ and\ \bibinfo {author} {\bibfnamefont {N.}~\bibnamefont {Gedik}},\ }\href {https://doi.org/10.1103/PhysRevB.86.035327} {\bibfield  {journal} {\bibinfo  {journal} {Phys. Rev. B}\ }\textbf {\bibinfo {volume} {86}},\ \bibinfo {pages} {035327} (\bibinfo {year} {2012})}\BibitemShut {NoStop}%
\bibitem [{\citenamefont {Boyd}(2008)}]{10.5555/1817101}%
  \BibitemOpen
  \bibfield  {author} {\bibinfo {author} {\bibfnamefont {R.~W.}\ \bibnamefont {Boyd}},\ }\href@noop {} {\emph {\bibinfo {title} {Nonlinear Optics, Third Edition}}},\ \bibinfo {edition} {3rd}\ ed.\ (\bibinfo  {publisher} {Academic Press, Inc.},\ \bibinfo {address} {USA},\ \bibinfo {year} {2008})\BibitemShut {NoStop}%
\bibitem [{\citenamefont {Butcher}\ and\ \citenamefont {Cotter}(1990)}]{butcher_cotter_1990}%
  \BibitemOpen
  \bibfield  {author} {\bibinfo {author} {\bibfnamefont {P.~N.}\ \bibnamefont {Butcher}}\ and\ \bibinfo {author} {\bibfnamefont {D.}~\bibnamefont {Cotter}},\ }\href {https://doi.org/10.1017/CBO9781139167994} {\emph {\bibinfo {title} {The Elements of Nonlinear Optics}}},\ Cambridge Studies in Modern Optics\ (\bibinfo  {publisher} {Cambridge University Press},\ \bibinfo {year} {1990})\BibitemShut {NoStop}%
\bibitem [{\citenamefont {Jonsson}(2003)}]{Jonsson25333}%
  \BibitemOpen
  \bibfield  {author} {\bibinfo {author} {\bibfnamefont {F.}~\bibnamefont {Jonsson}},\ }\href {http://jonsson.eu/research/lectures/nlo2003/compiled/nlo2003.pdf} {\emph {\bibinfo {title} {Lecture notes on nonlinear optics}}},\ \bibinfo {type} {Tech. Rep.}\ (\bibinfo  {institution} {KTH, Physics},\ \bibinfo {year} {2003})\ \bibinfo {note} {nR 20140804}\BibitemShut {NoStop}%
\bibitem [{\citenamefont {Shen}(2002)}]{ShenBook}%
  \BibitemOpen
  \bibfield  {author} {\bibinfo {author} {\bibfnamefont {Y.~R.}\ \bibnamefont {Shen}},\ }\href@noop {} {\emph {\bibinfo {title} {The Principles of Nonlinear Optics}}},\ \bibinfo {edition} {2nd}\ ed.\ (\bibinfo  {publisher} {John Wiley $\&$ Sons},\ \bibinfo {address} {New York},\ \bibinfo {year} {2002})\BibitemShut {NoStop}%
\bibitem [{\citenamefont {Bloembergen}(1996)}]{doi:10.1142/3046}%
  \BibitemOpen
  \bibfield  {author} {\bibinfo {author} {\bibfnamefont {N.}~\bibnamefont {Bloembergen}},\ }\href {https://doi.org/10.1142/3046} {\emph {\bibinfo {title} {Nonlinear Optics}}},\ \bibinfo {edition} {4th}\ ed.\ (\bibinfo  {publisher} {WORLD SCIENTIFIC},\ \bibinfo {year} {1996})\BibitemShut {NoStop}%
\bibitem [{\citenamefont {Birss}(1966)}]{0444101446}%
  \BibitemOpen
  \bibfield  {author} {\bibinfo {author} {\bibfnamefont {R.~R.}\ \bibnamefont {Birss}},\ }\href@noop {} {\emph {\bibinfo {title} {Symmetry and magnetism, second edition}}},\ \bibinfo {edition} {2nd}\ ed.\ (\bibinfo  {publisher} {North-Holland Publishing Company},\ \bibinfo {address} {USA},\ \bibinfo {year} {1966})\BibitemShut {NoStop}%
\bibitem [{\citenamefont {Gallego}\ \emph {et~al.}(2019)\citenamefont {Gallego}, \citenamefont {Etxebarria}, \citenamefont {Elcoro}, \citenamefont {Tasci},\ and\ \citenamefont {Perez-Mato}}]{Gallego:lk5043}%
  \BibitemOpen
  \bibfield  {author} {\bibinfo {author} {\bibfnamefont {S.~V.}\ \bibnamefont {Gallego}}, \bibinfo {author} {\bibfnamefont {J.}~\bibnamefont {Etxebarria}}, \bibinfo {author} {\bibfnamefont {L.}~\bibnamefont {Elcoro}}, \bibinfo {author} {\bibfnamefont {E.~S.}\ \bibnamefont {Tasci}},\ and\ \bibinfo {author} {\bibfnamefont {J.~M.}\ \bibnamefont {Perez-Mato}},\ }\href {https://doi.org/10.1107/S2053273319001748} {\bibfield  {journal} {\bibinfo  {journal} {Acta Crystallographica Section A}\ }\textbf {\bibinfo {volume} {75}},\ \bibinfo {pages} {438} (\bibinfo {year} {2019})}\BibitemShut {NoStop}%
\bibitem [{\citenamefont {Yamada}\ and\ \citenamefont {Kimoura}(1994)}]{PhysicalReviewBVol49No20}%
  \BibitemOpen
  \bibfield  {author} {\bibinfo {author} {\bibfnamefont {C.}~\bibnamefont {Yamada}}\ and\ \bibinfo {author} {\bibfnamefont {T.}~\bibnamefont {Kimoura}},\ }\href {https://doi.org/10.1103/PhysRevB.49.14372} {\bibfield  {journal} {\bibinfo  {journal} {Physical Review B}\ }\textbf {\bibinfo {volume} {49}},\ \bibinfo {pages} {14372} (\bibinfo {year} {1994})}\BibitemShut {NoStop}%
\bibitem [{\citenamefont {Li}\ \emph {et~al.}(2015)\citenamefont {Li}, \citenamefont {Su}, \citenamefont {Yang},\ and\ \citenamefont {Zhang}}]{PhysRevB.91.235128}%
  \BibitemOpen
  \bibfield  {author} {\bibinfo {author} {\bibfnamefont {Z.}~\bibnamefont {Li}}, \bibinfo {author} {\bibfnamefont {H.}~\bibnamefont {Su}}, \bibinfo {author} {\bibfnamefont {X.}~\bibnamefont {Yang}},\ and\ \bibinfo {author} {\bibfnamefont {J.}~\bibnamefont {Zhang}},\ }\href {https://doi.org/10.1103/PhysRevB.91.235128} {\bibfield  {journal} {\bibinfo  {journal} {Phys. Rev. B}\ }\textbf {\bibinfo {volume} {91}},\ \bibinfo {pages} {235128} (\bibinfo {year} {2015})}\BibitemShut {NoStop}%
\bibitem [{\citenamefont {Wosnitza}\ \emph {et~al.}(2006)\citenamefont {Wosnitza}, \citenamefont {Goll}, \citenamefont {Bianchi}, \citenamefont {Bergk}, \citenamefont {Kozlova}, \citenamefont {Opahle}, \citenamefont {Elgazzar}, \citenamefont {Richter}, \citenamefont {Stockert}, \citenamefont {v~L{\"o}hneysen}, \citenamefont {Yoshino},\ and\ \citenamefont {Takabatake}}]{Wosnitza_2006}%
  \BibitemOpen
  \bibfield  {author} {\bibinfo {author} {\bibfnamefont {J.}~\bibnamefont {Wosnitza}}, \bibinfo {author} {\bibfnamefont {G.}~\bibnamefont {Goll}}, \bibinfo {author} {\bibfnamefont {A.~D.}\ \bibnamefont {Bianchi}}, \bibinfo {author} {\bibfnamefont {B.}~\bibnamefont {Bergk}}, \bibinfo {author} {\bibfnamefont {N.}~\bibnamefont {Kozlova}}, \bibinfo {author} {\bibfnamefont {I.}~\bibnamefont {Opahle}}, \bibinfo {author} {\bibfnamefont {S.}~\bibnamefont {Elgazzar}}, \bibinfo {author} {\bibfnamefont {M.}~\bibnamefont {Richter}}, \bibinfo {author} {\bibfnamefont {O.}~\bibnamefont {Stockert}}, \bibinfo {author} {\bibfnamefont {H.}~\bibnamefont {v~L{\"o}hneysen}}, \bibinfo {author} {\bibfnamefont {T.}~\bibnamefont {Yoshino}},\ and\ \bibinfo {author} {\bibfnamefont {T.}~\bibnamefont {Takabatake}},\ }\href {https://doi.org/10.1088/1367-2630/8/9/174} {\bibfield  {journal} {\bibinfo  {journal} {New Journal of Physics}\ }\textbf {\bibinfo {volume} {8}},\ \bibinfo {pages} {174} (\bibinfo {year} {2006})}\BibitemShut {NoStop}%
\bibitem [{\citenamefont {Fiebig}\ \emph {et~al.}(2005)\citenamefont {Fiebig}, \citenamefont {Pavlov},\ and\ \citenamefont {Pisarev}}]{Fiebig:05}%
  \BibitemOpen
  \bibfield  {author} {\bibinfo {author} {\bibfnamefont {M.}~\bibnamefont {Fiebig}}, \bibinfo {author} {\bibfnamefont {V.~V.}\ \bibnamefont {Pavlov}},\ and\ \bibinfo {author} {\bibfnamefont {R.~V.}\ \bibnamefont {Pisarev}},\ }\href {https://doi.org/10.1364/JOSAB.22.000096} {\bibfield  {journal} {\bibinfo  {journal} {J. Opt. Soc. Am. B}\ }\textbf {\bibinfo {volume} {22}},\ \bibinfo {pages} {96} (\bibinfo {year} {2005})}\BibitemShut {NoStop}%
\bibitem [{\citenamefont {Ahn}\ \emph {et~al.}(2024)\citenamefont {Ahn}, \citenamefont {Guo}, \citenamefont {Xue}, \citenamefont {Qu}, \citenamefont {Sun}, \citenamefont {Mandrus},\ and\ \citenamefont {Zhao}}]{Ahn2024}%
  \BibitemOpen
  \bibfield  {author} {\bibinfo {author} {\bibfnamefont {Y.}~\bibnamefont {Ahn}}, \bibinfo {author} {\bibfnamefont {X.}~\bibnamefont {Guo}}, \bibinfo {author} {\bibfnamefont {R.}~\bibnamefont {Xue}}, \bibinfo {author} {\bibfnamefont {K.}~\bibnamefont {Qu}}, \bibinfo {author} {\bibfnamefont {K.}~\bibnamefont {Sun}}, \bibinfo {author} {\bibfnamefont {D.}~\bibnamefont {Mandrus}},\ and\ \bibinfo {author} {\bibfnamefont {L.}~\bibnamefont {Zhao}},\ }\href {https://doi.org/10.1038/s41566-023-01300-2} {\bibfield  {journal} {\bibinfo  {journal} {Nature Photonics}\ }\textbf {\bibinfo {volume} {18}},\ \bibinfo {pages} {26} (\bibinfo {year} {2024})}\BibitemShut {NoStop}%
\bibitem [{\citenamefont {M\"uller}\ \emph {et~al.}(2015)\citenamefont {M\"uller}, \citenamefont {Desilets-Benoit}, \citenamefont {Gauthier}, \citenamefont {Lapointe}, \citenamefont {Bianchi}, \citenamefont {Maris}, \citenamefont {Zahn}, \citenamefont {Beyer}, \citenamefont {Green}, \citenamefont {Wosnitza}, \citenamefont {Yamani},\ and\ \citenamefont {Kenzelmann}}]{PhysRevB.92.184432}%
  \BibitemOpen
  \bibfield  {author} {\bibinfo {author} {\bibfnamefont {R.~A.}\ \bibnamefont {M\"uller}}, \bibinfo {author} {\bibfnamefont {A.}~\bibnamefont {Desilets-Benoit}}, \bibinfo {author} {\bibfnamefont {N.}~\bibnamefont {Gauthier}}, \bibinfo {author} {\bibfnamefont {L.}~\bibnamefont {Lapointe}}, \bibinfo {author} {\bibfnamefont {A.~D.}\ \bibnamefont {Bianchi}}, \bibinfo {author} {\bibfnamefont {T.}~\bibnamefont {Maris}}, \bibinfo {author} {\bibfnamefont {R.}~\bibnamefont {Zahn}}, \bibinfo {author} {\bibfnamefont {R.}~\bibnamefont {Beyer}}, \bibinfo {author} {\bibfnamefont {E.}~\bibnamefont {Green}}, \bibinfo {author} {\bibfnamefont {J.}~\bibnamefont {Wosnitza}}, \bibinfo {author} {\bibfnamefont {Z.}~\bibnamefont {Yamani}},\ and\ \bibinfo {author} {\bibfnamefont {M.}~\bibnamefont {Kenzelmann}},\ }\href {https://doi.org/10.1103/PhysRevB.92.184432} {\bibfield  {journal} {\bibinfo  {journal} {Phys. Rev. B}\ }\textbf {\bibinfo {volume} {92}},\ \bibinfo {pages} {184432} (\bibinfo {year} {2015})}\BibitemShut {NoStop}%
\bibitem [{\citenamefont {Wosnitza}(2007)}]{Wosnitza2007}%
  \BibitemOpen
  \bibfield  {author} {\bibinfo {author} {\bibfnamefont {J.}~\bibnamefont {Wosnitza}},\ }\href {https://doi.org/10.1007/s10909-007-9309-x} {\bibfield  {journal} {\bibinfo  {journal} {Journal of Low Temperature Physics}\ }\textbf {\bibinfo {volume} {147}},\ \bibinfo {pages} {249} (\bibinfo {year} {2007})}\BibitemShut {NoStop}%
\bibitem [{\citenamefont {Kreyssig}\ \emph {et~al.}(2011)\citenamefont {Kreyssig}, \citenamefont {Kim}, \citenamefont {Kim}, \citenamefont {Pratt}, \citenamefont {Sauerbrei}, \citenamefont {March}, \citenamefont {Tesdall}, \citenamefont {Bud'ko}, \citenamefont {Canfield}, \citenamefont {McQueeney},\ and\ \citenamefont {Goldman}}]{PhysRevB.84.220408}%
  \BibitemOpen
  \bibfield  {author} {\bibinfo {author} {\bibfnamefont {A.}~\bibnamefont {Kreyssig}}, \bibinfo {author} {\bibfnamefont {M.~G.}\ \bibnamefont {Kim}}, \bibinfo {author} {\bibfnamefont {J.~W.}\ \bibnamefont {Kim}}, \bibinfo {author} {\bibfnamefont {D.~K.}\ \bibnamefont {Pratt}}, \bibinfo {author} {\bibfnamefont {S.~M.}\ \bibnamefont {Sauerbrei}}, \bibinfo {author} {\bibfnamefont {S.~D.}\ \bibnamefont {March}}, \bibinfo {author} {\bibfnamefont {G.~R.}\ \bibnamefont {Tesdall}}, \bibinfo {author} {\bibfnamefont {S.~L.}\ \bibnamefont {Bud'ko}}, \bibinfo {author} {\bibfnamefont {P.~C.}\ \bibnamefont {Canfield}}, \bibinfo {author} {\bibfnamefont {R.~J.}\ \bibnamefont {McQueeney}},\ and\ \bibinfo {author} {\bibfnamefont {A.~I.}\ \bibnamefont {Goldman}},\ }\href {https://doi.org/10.1103/PhysRevB.84.220408} {\bibfield  {journal} {\bibinfo  {journal} {Phys. Rev. B}\ }\textbf {\bibinfo {volume} {84}},\ \bibinfo {pages} {220408} (\bibinfo {year} {2011})}\BibitemShut {NoStop}%
\bibitem [{\citenamefont {Sa}\ \emph {et~al.}(2000)\citenamefont {Sa}, \citenamefont {Valent{\'\i}},\ and\ \citenamefont {Gros}}]{Sa2000}%
  \BibitemOpen
  \bibfield  {author} {\bibinfo {author} {\bibfnamefont {D.}~\bibnamefont {Sa}}, \bibinfo {author} {\bibfnamefont {R.}~\bibnamefont {Valent{\'\i}}},\ and\ \bibinfo {author} {\bibfnamefont {C.}~\bibnamefont {Gros}},\ }\href {https://doi.org/10.1007/s100510050133} {\bibfield  {journal} {\bibinfo  {journal} {The European Physical Journal B - Condensed Matter and Complex Systems}\ }\textbf {\bibinfo {volume} {14}},\ \bibinfo {pages} {301} (\bibinfo {year} {2000})}\BibitemShut {NoStop}%
\bibitem [{\citenamefont {Muthukumar}\ \emph {et~al.}(1996)\citenamefont {Muthukumar}, \citenamefont {Valent\'{\i}},\ and\ \citenamefont {Gros}}]{PhysRevB.54.433}%
  \BibitemOpen
  \bibfield  {author} {\bibinfo {author} {\bibfnamefont {V.~N.}\ \bibnamefont {Muthukumar}}, \bibinfo {author} {\bibfnamefont {R.}~\bibnamefont {Valent\'{\i}}},\ and\ \bibinfo {author} {\bibfnamefont {C.}~\bibnamefont {Gros}},\ }\href {https://doi.org/10.1103/PhysRevB.54.433} {\bibfield  {journal} {\bibinfo  {journal} {Phys. Rev. B}\ }\textbf {\bibinfo {volume} {54}},\ \bibinfo {pages} {433} (\bibinfo {year} {1996})}\BibitemShut {NoStop}%
\bibitem [{\citenamefont {Muto}\ \emph {et~al.}(1998)\citenamefont {Muto}, \citenamefont {Tanabe}, \citenamefont {Iizuka-Sakano},\ and\ \citenamefont {Hanamura}}]{PhysRevB.57.9586}%
  \BibitemOpen
  \bibfield  {author} {\bibinfo {author} {\bibfnamefont {M.}~\bibnamefont {Muto}}, \bibinfo {author} {\bibfnamefont {Y.}~\bibnamefont {Tanabe}}, \bibinfo {author} {\bibfnamefont {T.}~\bibnamefont {Iizuka-Sakano}},\ and\ \bibinfo {author} {\bibfnamefont {E.}~\bibnamefont {Hanamura}},\ }\href {https://doi.org/10.1103/PhysRevB.57.9586} {\bibfield  {journal} {\bibinfo  {journal} {Phys. Rev. B}\ }\textbf {\bibinfo {volume} {57}},\ \bibinfo {pages} {9586} (\bibinfo {year} {1998})}\BibitemShut {NoStop}%
\bibitem [{\citenamefont {Hirschberger}\ \emph {et~al.}(2016{\natexlab{b}})\citenamefont {Hirschberger}, \citenamefont {Kushwaha}, \citenamefont {Wang}, \citenamefont {Gibson}, \citenamefont {Liang}, \citenamefont {Belvin}, \citenamefont {Bernevig}, \citenamefont {Cava},\ and\ \citenamefont {Ong}}]{Hirschberger2016}%
  \BibitemOpen
  \bibfield  {author} {\bibinfo {author} {\bibfnamefont {M.}~\bibnamefont {Hirschberger}}, \bibinfo {author} {\bibfnamefont {S.}~\bibnamefont {Kushwaha}}, \bibinfo {author} {\bibfnamefont {Z.}~\bibnamefont {Wang}}, \bibinfo {author} {\bibfnamefont {Q.}~\bibnamefont {Gibson}}, \bibinfo {author} {\bibfnamefont {S.}~\bibnamefont {Liang}}, \bibinfo {author} {\bibfnamefont {C.~A.}\ \bibnamefont {Belvin}}, \bibinfo {author} {\bibfnamefont {B.~A.}\ \bibnamefont {Bernevig}}, \bibinfo {author} {\bibfnamefont {R.~J.}\ \bibnamefont {Cava}},\ and\ \bibinfo {author} {\bibfnamefont {N.~P.}\ \bibnamefont {Ong}},\ }\href {https://doi.org/10.1038/nmat4684} {\bibfield  {journal} {\bibinfo  {journal} {Nature Materials}\ }\textbf {\bibinfo {volume} {15}},\ \bibinfo {pages} {1161} (\bibinfo {year} {2016}{\natexlab{b}})}\BibitemShut {NoStop}%
\end{thebibliography}
\end{document}